# Hardware-Efficient Photonic Tensor Core: Accelerating Deep Neural Networks with Structured Compression

Shupeng Ning,[a] Hanqing Zhu,[a] Chenghao Feng,[a] Jiaqi Gu,[b] David Z. Pan,[a] and Ray T. Chen[a,*]

[a] *Department of Electrical and Computer Engineering, The University of Texas at Austin, Austin, Texas 78758, United States*

[b] *School of Electrical, Computer and Energy Engineering, Arizona State University, Tempe, Arizona 85281, United States*

[*]*Email*: chenrt@austin.utexas.edu

## Abstract

The rapid growth in computing demands, particularly driven by artificial intelligence (AI) applications, has begun to exceed the capabilities of traditional electronic hardware. Optical computing offers a promising alternative due to its parallelism, high computational speed, and low power consumption. However, existing photonic integrated circuits (PICs) are constrained by large footprints, costly electro-optical (E-O) interfaces, and complex control mechanisms, limiting the practical scalability of optical neural networks (ONNs). To address these limitations, we introduce a block-circulant photonic tensor core (CirPTC) for a structure-compressed optical neural network (StrC-ONN) architecture. The structured compression technique substantially reduces both model complexity and hardware resources without sacrificing the versatility of neural networks, and achieves accuracy comparable to uncompressed models. Additionally, we propose a hardware-aware training framework to compensate for on-chip nonidealities to improve model robustness and accuracy. Experimental validation through image processing and classification tasks demonstrates that our StrC-ONN achieves a reduction in trainable parameters of up to 74.91%, while still maintaining competitive accuracy levels. Performance analyses further indicate that this hardware–software co-design approach is expected to yield a 3.56× improvement in power efficiency. By reducing both hardware requirements and control complexity across multiple dimensions, this work explores a new pathway toward practical and scalable ONNs, highlighting a promising route to address future computational efficiency challenges.

## Introduction

Machine learning (ML) with DNNs has transformed various aspects of science and technology,[1] including object recognition,[2]-[4] autonomous driving,[5] natural language processing,[6][7] and medical diagnosis.[8][9] Additionally, the emergence of large language models (LLMs) has further demonstrated human-level intelligence in specific tasks.[10][11] The unprecedented advancements of modern DNNs are driven by rapidly expanding model sizes—with millions to tens of billions of parameters—and increasing data volumes, which allow for the extraction of intricate, high-level features needed for complex tasks.[1][11][12] Notably, these improvements in model performance have led to a surge in demand for computing resources and memory access. DNNs typically comprise multiple cascaded layers, where data are represented as vectors processed through matrix-vector multiplications (MVMs) with corresponding weights, which are the main contributors to time and power consumption. To efficiently execute MVMs, hardware AI accelerators—including, but not limited to, graphical processing units (GPUs), field-programmable gate arrays (FPGAs),[13][14] and application-specific integrated circuits (ASICs)[15]—have been widely developed by both industry and academia. However, as integrated circuits scale to include tens of billions of transistors operating at gigahertz (GHz), they encounter saturated power efficiency, heat dissipation issues, and limited bandwidth, making electrical processors unsustainable for meeting surging demands.[16] Moreover, as semiconductor processes advance to sub-nanometer nodes and approach the inherent physical limitations of devices, the unreliability caused by quantum uncertainties has become another bottleneck for further scaling.[16][17] Pursuing higher computing density, power efficiency, and scalability remains a persistent goal in developing high-performance AI accelerators.[18]

Due to these intrinsic bottlenecks, the exploration of novel technologies beyond traditional electrical digital computing has emerged as an attractive trend.[19]-[24] Among these, optical neural networks (ONNs) based on PICs are promising candidates for AI accelerators. The inherent high computational speeds, low power consumption, low latency, and high parallelism enabled by the unique multiplexing techniques of optical computing can effectively overcome the aforementioned issues.[25] Additionally, advances in silicon photonics allow for the implementation of optical computing on low-cost PICs with high integration density using CMOS-compatible fabrication processes. Over the past decade, various PIC-based ONN prototypes have been presented, demonstrating the implementation of multilayer perceptrons (MLPs),[22][23] convolutional neural networks (CNNs),[24][26][27] spike neural networks (SNNs),[28] *etc*. However, a primary challenge in contemporary PICs arises from the micron- to millimeter-scale dimensions of on-chip photonic devices, which leads to unavoidable trade-offs between the PIC scaling and the chip footprint.[25] Furthermore, the peripheral electrical components for E-O modulation, data access, and conversion between analog and digital domains boost power consumption, thereby undermining the power efficiency benefits of optical computing.

To address these challenges associated with implementing large-scale MVMs on PICs, it is necessary to explore strategies that extend beyond device- or circuit-level innovations. One critical strategy is domain-specific hardware customization informed by algorithm-level insights, which can significantly improve the efficiency and scalability of photonic tensor cores (PTCs). Recent research indicates that DNNs, especially CNNs, are often over-parameterized with significant redundancy in their parameters.[19][29] This redundancy leads to unnecessary power consumption, prolonged runtimes, and increased memory usage, which has driven extensive research into algorithm-level model compression techniques such as connection pruning,[30][31] low-rank approximations,[32][33] and structured model designs.[19][34] These approaches have demonstrated substantial improvements in hardware efficiency with minimal performance trade-offs, making them highly relevant for optimizing optical computing systems. While several ONNs employing compression techniques have been proposed, these implementations, which use free-space optics or conventional mesh structures,[35][36] typically have complex systems or large footprints with significant challenges for calibration and precise programming. These factors collectively restrict their ability to fully leverage the efficiency benefits that novel algorithms might offer.

In this work, we advance the exploration of model compression techniques for optical computing by introducing a customized StrC-ONN architecture with the CirPTC. On the algorithmic side, StrC-ONN represents weight matrices using block-circulant matrices (BCMs) with restricted parameter spaces, effectively eliminating redundant parameters and unlocking the potential for designing customized ONN structures to fully harness their efficiency. From the hardware perspective, a compact crossbar array is specifically designed to implement BCMs, thereby minimizing E-O interface costs while achieving high area and power efficiency. Notably, CirPTC imposes structured compression directly through its circuit topology, and the cascading of its building blocks enables a one-shot calibration mechanism. This feature streamlines device characterization, whereas most existing ONN systems still rely on labor-intensive iterative calibration of modulators. To address on-chip nonidealities, we propose and experimentally deploy a hardware-aware training framework that compensates for these imperfections, enhancing model performance and robustness. We experimentally demonstrate the image processing capabilities of our design using convolutional kernels. Furthermore, we evaluate the end-to-end accuracy of StrC-ONN on various image classification tasks, including the SVHN, CIFAR-10, and COVID-QU-Ex datasets. This approach achieves up to a 74.91 % reduction in trainable parameters, active modulators and memory requirements, while maintaining comparable accuracy to conventional general matrix multiplication (GEMM)-based digital DNNs. Performance analysis reveals that the design can achieve a computational density of 5.84 tera operations per second (TOPS) per $mm^2$, with a power efficiency of 17.13 TOPS/W—representing a 3.56× improvement enabled by the dedicated hardware-software co-design strategy—after appropriate scaling and a unique spectral folding technique. Therefore, the hardware–software co-designed architecture and CirPTC provide a viable pathway toward the practical deployment of optical computing systems, enhancing efficiency and practicability while laying the groundwork for next-generation AI hardware.

## Operation Principle

### ONN with Structured Compression

The structured compression, compared to other compression techniques, reduces both computational complexity and storage complexity while maintaining a regular network connection topology. An $M \times N$ BCM, for instance, is composed of $P \times Q$ blocks, each order-$l$ square matrix following the circulant format. As illustrated in Eq (1), the first-row vector (primary vector) in a circulant matrix $\boldsymbol{w_{ij}} = [w_{1,ij}, w_{2,ij}, \ldots, w_{l,ij}]$ contains all independent parameters, with subsequent rows being circulant reformations of it. Intuitively, compared to general matrices, block-circulant matrices reduce the number of independent parameters to $MN/l$.

$$\boldsymbol{W_{block}} = \overbrace{\begin{bmatrix} \boldsymbol{W_{11}} & \boldsymbol{W_{12}} & \cdots & \boldsymbol{W_{1q}} \\ \boldsymbol{W_{21}} & \boldsymbol{W_{22}} & \cdots & \boldsymbol{W_{2q}} \\ \vdots & \vdots & \boldsymbol{W_{ij}} & \vdots \\ \boldsymbol{W_{p1}} & \boldsymbol{W_{p2}} & \cdots & \boldsymbol{W_{pq}} \end{bmatrix}}^{N = Q \times l} \Bigg\} M = P \times l \quad ; \boldsymbol{W_{ij}} = \begin{bmatrix} w_{1,ij} & w_{2,ij} & \cdots & w_{l,ij} \\ w_{l,ij} & w_{1,ij} & \cdots & w_{l-1,ij} \\ \vdots & \vdots & \ddots & \vdots \\ w_{2,ij} & w_{3,ij} & \cdots & w_{1,ij} \end{bmatrix} \tag{1}$$

In DNNs, the mapping from an input vector $\boldsymbol{x} \in \mathbb{R}^N$ to an output vector $\boldsymbol{y} \in \mathbb{R}^M$ across two successive layers is typically formalized as $\boldsymbol{y} = \sigma(\boldsymbol{W} \cdot \boldsymbol{x} + \boldsymbol{b})$, where $\boldsymbol{W} \in \mathbb{R}^{M \times N}$ is the weight matrix, $\boldsymbol{b} \in \mathbb{R}^M$ is a bias vector and $\sigma(\cdot)$ is an element-wise nonlinear activation function. Crucially, regardless of whether the layer is fully connected, convolutional, recurrent or attention-based, its core computation can be reformulated as one or more MVMs via appropriate matrixization techniques. Therefore, the block-circulant compression technique applies universally to the weight matrices of diverse architectures. It is essential to recognize that while the StrC-ONN share a similar connection topology with conventional DNNs, there is no direct correspondence or conversion between the two architectures. During training, each weight matrix is constrained to a block-circulant format: an $M \times N$ matrix is partitioned into circulant blocks of size $l$, and the learning task reduces to identifying the length-$l$ vector that uniquely defines each circulant block. Owing to this characteristic, the StrC-ONN could improve efficiency across multiple dimensions.

Firstly, the compression strategy significantly reduces the number of model parameters and memory usage. During offline training, learnable tensor is instantiated as a set of primary vectors, and autograd accumulates gradients only for $P \times Q \times l$ explicit elements during backward propagation, rather than for every elements in the full weight matrix. From a hardware perspective, it decreases the number of active E-O modulators required for weight programming and the control complexity for on-chip inference. Additionally, this strategy reduces memory consumption for weight storage and decreases data access demands on hardware resources, such as digital-to-analog converters (DACs) and associated transmitter circuitry. However, these advantages come with restricted parameter space, presenting a potential trade-off. For instance, a small block size in the BCM yields a lower compression ratio, while a larger size offers substantial compression but may result in accuracy degradation. Related research has demonstrated the mathematical rigor of this approach.[37] Specifically, structured networks preserve the universal representability of DNNs, allowing them to approximate or represent functions with complexity comparable to those handled by uncompressed networks. Additionally, with appropriate compression, structured networks maintain comparable performance and accuracies across a wide range of tasks, which has been demonstrated in electrical digital computing.[19] In this work, we implemented the compression technique through a compact PIC design that achieves high hardware and power efficiency, and experimentally demonstrated its performance.

## Operation Mechanism of CirPTC

The operation mechanism of CirPTC and the ONN training framework are shown in Fig. 1. The primary vector $\boldsymbol{w}$ is encoded into incoherent light intensity using serial microring resonators (MRRs) operating at different wavelengths, which then physically multiply with the input vector $\boldsymbol{x}$ encoded by Mach–Zehnder modulators (MZMs).The CirPTC is characterized by an crossbar switch array in which the switches operate at different wavelengths following a block-circulant arrangement [Fig. 1(b)]. The switch array maps the elements of a weighted vector to the outputs, thereby directly implementing the structured configuration by the circuit topology. By leveraging the wavelength division multiplexing (WDM), on-chip photodetectors (PDs) can autonomously sum weighted elements at the output ports. Compared to ONN architectures designed for GEMMs, the CirPTC requires only $M \times N/l$ active MRRs to implement a $M \times N$ BCM, while the optical switches in the crossbar array are "static" or even passive (see details in Supplementary Note 8).[24] From a hardware perspective, this configuration significantly reduces reprogramming complexity and the number of DACs required for weight encoding.

As mentioned, the compression mechanism requires partitioning 2-D matrices and embedding structured constraints. The operation for the FC layers is straightforward because the connection between the two layers can be directly represented by a weight matrix. In convolutional layers, learnable kernels slide over the input data, generating feature maps through convolution operations. Unlike FC layers, convolution operations share kernels and focus on local areas, resulting in inherently sparse connections. To perform convolution operations efficiently, we utilize the "*im2col*" matrixization technique, which tiles all kernels and inputs into large 2-D matrices, transforming tensor-based convolution operations into matrix-matrix multiplications.[38][39] As illustrated in Fig. 1(a), each kernel comprises of $C_{in}$ channels with a dimension of $k \times k \times C_{in}$, and each set of kernels corresponds to an output feature map. For a convolution layer that includes $C_{out}$ output features, the kernels can be reorganized into a 2-D weight matrix $\boldsymbol{W}$ of size $C_{out} \times (k \times k \times C_{in})$ by flattening kernels into a 1-D vector and stacking them row-wise. Here, we constrain $\boldsymbol{W}$ to a block-circulant format starting from the training stage, and the primary vector $\boldsymbol{w_{ij}}$ for each circulant block are encoded by $N/l$ rails of serial MRRs. Similarly, as the kernel slides across the input image with $w \times h$ pixels, the corresponding pixels are rearranged into a column vector of dimension $(k \times k \times C_{in})$ and then stacked to create the input matrix $\boldsymbol{X}$. To perform MVMs with the circulant block, the column vectors in $\boldsymbol{X}$ are partitioned into length-$l$ subgroups. Based on this approach, the convolution operation for one feature map translates into $(w - k + 1)\cdot(h - k + 1)$ MVMs between a BCM and the input vectors of length $k^2 \times C_{in}$.

## Training Framework for CirPTC-Based ONN

As an analog computing platform, CirPTC inherently exhibits various hardware-induced nonidealities, including truncated resolution, crosstalk, and fabrication variances. Direct deployment of ML models faces challenges in accurately capturing complex chip behaviors, potentially leading to significant performance degradation. To address these issues, implementing DNNs on CirPTC with enhanced robustness requires a hardware-aware training strategy based on actual measurements. However, training models directly using on-chip optical responses presents an additional challenge, as the data behaves like a lookup table (LUT), which is inherently non-differentiable. In this work, we employ a hardware-aware training framework that incorporates a differentiable PIC estimator (DPE). Furthermore, we embed an adaptive coupling matrix that counteracts the coherent crosstalk within the crossbar array, thereby preserving model accuracy. The general procedures of the training framework are summarized in Fig. 1(c), with additional details provided in Supplementary Note 7. Following quantization and dynamic noise injection, the DPE operates in two modes: differentiable and lookup. We exhaustively sweep a large set of vector combinations to construct a LUT, then fit a surrogate model that approximates the on-chip forward pass. This surrogate captures the dominant non-idealities while remaining fully differentiable for back-propagation. Experimental results demonstrate that this approach effectively compensates for on-chip nonidealities, resulting in improved inference accuracy compared to previous ONN training protocols, such as simulation-based gradient approximation,[22][40] and derivative-free optimization.[41][42] Beyond the off-chip training, fully forward in-situ training of ONNs has been demonstrated in several pioneering works, showcasing the potential of updating network parameters directly on-chip.[43][44] While these approaches reduce modeling errors effectively, it comes at the cost of increased system complexity, power consumption, and chip area usage for integrated monitors and real-time feedback loops. In contrast, our hardware-aware offline training embeds actual device nonidealities into the training loop, enhancing robustness to chip-level variations without extra hardware overhead. This streamlined strategy offers a practical, scalable solution for ONN deployment. Moreover, the further combination of on-chip training protocols and accepting the required hardware investment could further improve overall accuracy.

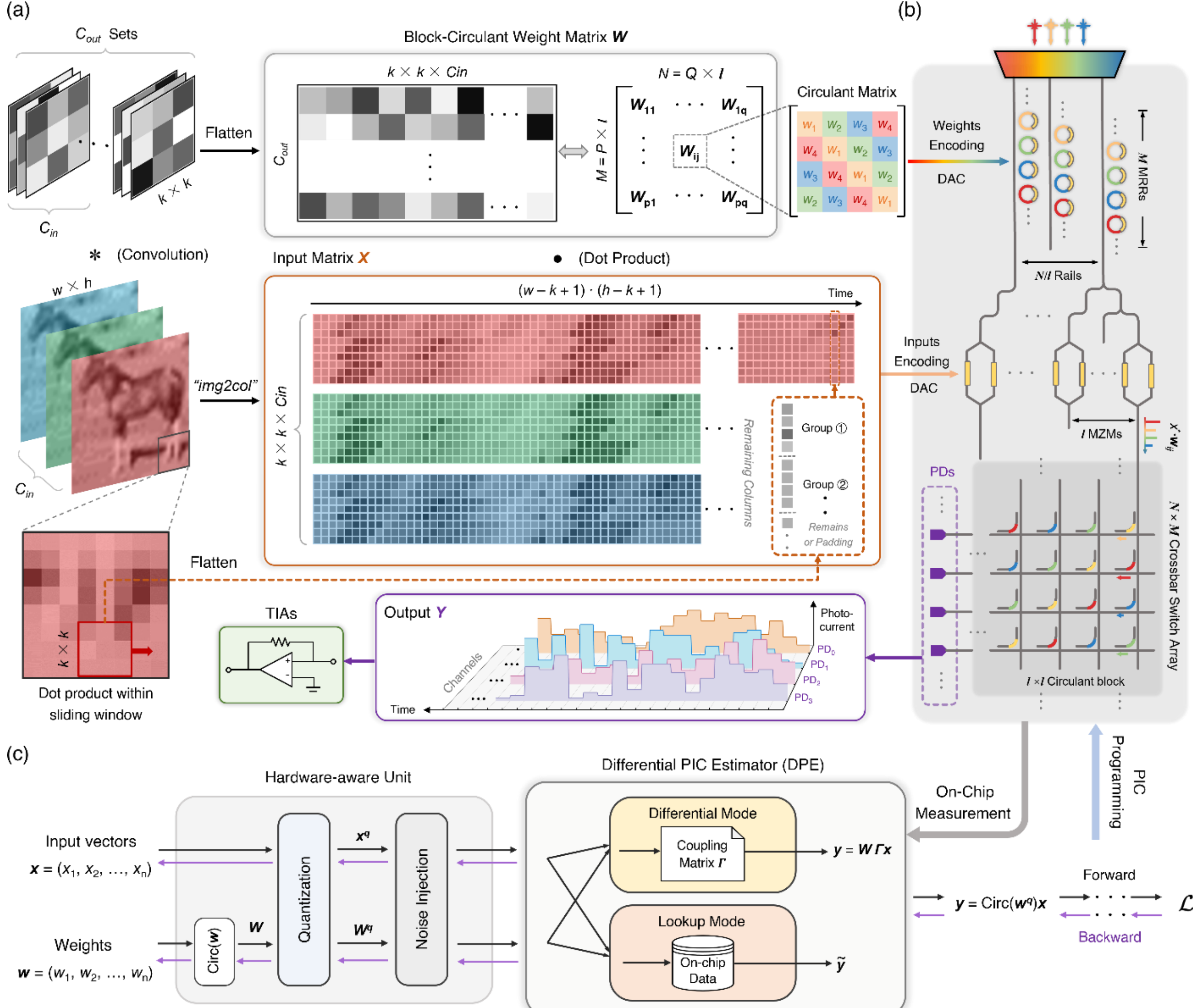

**Fig. 1 General architecture of StrC-ONN and image convolution operations using CirPTC.** (a) Transformation of convolution operations into MVMs and the implementation of BCM on CirPTC. This conceptual illustration shows an input image with 3 channels perform convolution with $C_{out}$ sets of $k \times k \times C_{in}$ kernels (where $k$ =3). The 2-D flattened weight matrix $\boldsymbol{W}$ is configured as a BCM with 4×4 circulant blocks. Pixels within the sliding window are partitioned into subgroups of length 4 after flattening and are then sent to the CirPTC. (b) General schematic of order-$l$ CirPTC with a $N \times M$ crossbar array. The output photocurrent is detected by oscilloscope or ADC after amplification by the off-chip trans-impedance amplifier (TIA). Here, different colors denote devices operating at different wavelengths (c) The hardware-aware training framework for CirPTC-based ONN.

## Result

### Design of CirPTC

In this work, we designed and fabricated an order-4 CirPTC, selecting this order as the optimal compromise between DNN pre-training accuracy across different block sizes and overall chip footprint (see details in Supplementary Note 7). The micrograph of the chip after packaging is shown in Fig. 2(a), with the zoom-in figures of its key components [Fig. 2(b)]. A continuous-wave (CW) multiwavelength input is coupled to the chip via an edge coupler. To prevent crosstalk, the operational ranges of 4 MRRs are set to be separated without overlapping in the spectrum, as illustrated in Fig. 2(d). To enable hardware-efficient scalar multiplication $x^* \cdot \boldsymbol{w_{ij}}$ ($x^*$ is one element in $\boldsymbol{x_j}$) within a single device, the modulator for input encoding needs to modulate signals across multiple wavelengths simultaneously [Fig. 2(e)]. This requires modulators with broadband transmission characteristics, like those provided by MZMs based on the phase-tuning mechanism.[25][45] For the crossbar switch array, we employed 16 add-drop MRRs, each interconnected with others through shared input and drop bus waveguides along the same row and column. Each MRR is calibrated to a designated wavelength according to the circulant configuration, thereby redirecting the appropriate element from $x^* \cdot \boldsymbol{w_{ij}}$ to the PDs. Finally, the PDs automatically sum the signals on the column bus waveguide as photocurrent, thereby completing the MVM operation. To avoid distortion of the circulant block, each MRR needs to be calibrated to achieve a uniform maximum output [grey dotted line in Fig. 2(f)]. The primary advantage of the crossbar array is the small footprint of MRR and the inherent sharing of waveguides, both of which enhance the compactness and scalability. Since these MRRs function exclusively as wavelength-dependent filters, their operating statuses are fixed after calibration. Additionally, cascading each building block enables a one-shot calibration mechanism that minimizes the impact of dynamic nonidealities, such as thermal crosstalk and loss, while simplifying control complexity. Therefore, CirPTC proposes an efficient, customized PIC design through a hardware-software co-design approach, rather than merely exploring algorithmic characteristics in isolation. Further details about system calibration are provided in Supplementary Note 1.

For yield and cost considerations, the CirPTC utilizes thermo-optic modulators from the foundry's Process Design Kit (PDK), with each device type having identical specifications. The ohmic microheaters of MZMs and MRRs are programmed by a multi-channel DAC. Since all MRRs are identical (exhibiting similar resonant wavelengths at zero bias), the four operating wavelengths selected—1545.5, 1551.0, 1560.5, and 1563.0 nm—are spaced separately within a single free spectral range (FSR) to minimize spectral crosstalk [Fig. 2(d)&(f)]. In future optimizations, the radius of MRRs can be customized to achieve configurations with different resonant wavelengths, thereby further decreasing power consumption. The on-chip photodetectors, along with off-chip TIAs, convert the optical output to electrical voltage signals, which are then measured by oscilloscopes. The CirPTC is mounted on a customized printed circuit board (PCB) and connected to the control units via wire bonding [Fig. 2(c)]. To facilitate calibration and monitoring, four monitor ports at the ends of horizontal bus waveguides are coupled to the fiber array. Due to the cascading and independence of the building blocks, we can measure the transmission characteristic of each device through the on-chip photodetector or the monitor port. Based on measurements and physical models of devices, we fitted the transmission curves of MZMs and MRRs for subsequent demonstrations and experiments [Fig. 2(d)-(f)].

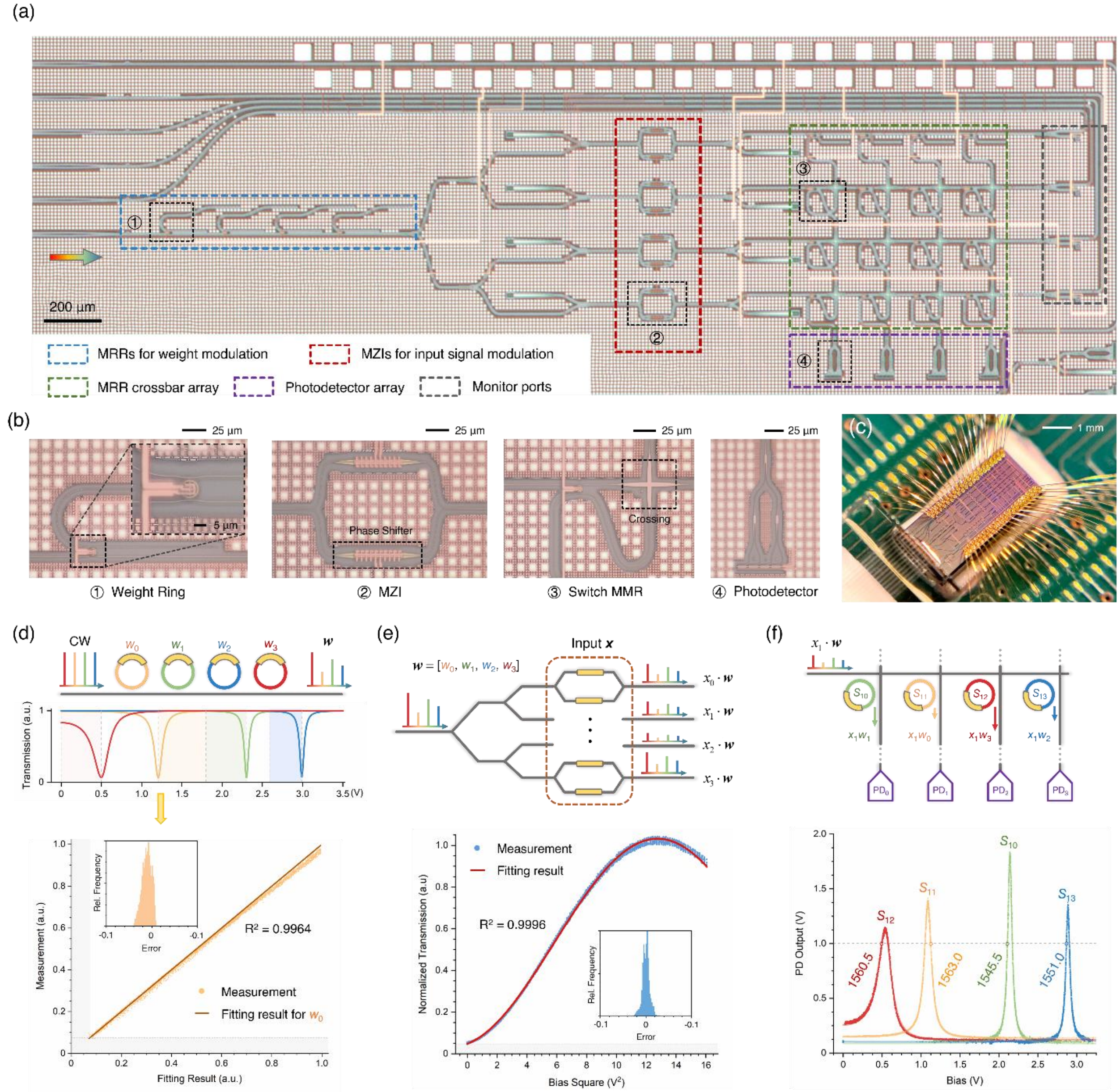

**Fig. 2 Schematic of an order-4 CirPTC.** (a) Optical micrograph of the CirPTC, featuring five main building blocks. (b) Key optical components highlighted in a zoomed-in micrograph. (c) Electrical and optical package for the CirPTC. (d)-(f) Illustrations of the operational mechanisms and data flow within each building block of CirPTC. Notably, to avoid spectral overlap, the modulation range for MRRs operating at 1563.0 nm is allocated to the right half-branch of resonate peak, whereas the modulation ranges for the other MRRs are situated on the left branch. The on-chip measurement results (dots in the background) and the fitting results based on physical models are shown in the lower part of figures. Due to the dark current of the photodetector and the asymmetric, lossy coupling of the MRRs,[46] a fixed “forbidden zone” (grey area) is established at each output port. However, this can be eliminated through post-processing in the electrical domain. The transmission characteristics of other devices are provided in Supplementary Note 1.

## On-chip image processing

To experimentally illustrate the aforementioned principle, we performed on-chip image processing with convolutional kernels that extract physical features from input images, thereby evaluating both CirPTC functionality and on-chip MVM accuracy. Solely for demonstration purposes, we implemented several kernels with clear intuitive interpretations. Although CirPTC enforces a block-circulant structure on the weight matrix, arbitrary kernels can still be realized by mapping each *im2col*-flatten kernel vector to a single column of BCM and then extending with block-circulant format (see Supplementary Note 5). We should note that this procedure confines all meaningful weights to the designated column, with the remaining columns serving only as by-products of the BCM representation. Firstly, we demonstrate the convolution operation on input images from the CIFAR-10 dataset using a 3×3 blur kernel [Fig. 3(a)], which results in a 12 × 4 BCM with an addition of 3 rows of padding. Given that the three channels (RGB) are convolved with the same 2D blur kernel, they can be reorganized into an input matrix of $k^2 \times 3\cdot(n-k+1)^2$, where $k = 3$ and $n = 32$. Then, the input matrix $\boldsymbol{X}$ encoded by an FPGA with a 4-bit resolution and a time interval of $\tau = 80$ μs, corresponding to a data rate of 12.5 Kbaud [Fig. 3(b)]. Given the extensive weight sharing in CNNs, the bottleneck of data rate primarily arises from the limited output settling time of the DAC (20 μs) and the input encoding bandwidth in the tens of KHz range. In future optimizations, the throughput of CirPTC can be increased to GHz levels by utilizing high-speed DACs and E-O modulators, such as carrier-depletion and carrier-accumulation MZMs.[25][47] Fig. 3(a) presents the on-chip convolution results of four images (see Supplementary Note 5 for results of all images), and Fig. 3(c) displays the experimental waveform alongside the expected values (grey lines) for the “horse”. Each time slot in the waveform results from the post-processing summation of the dot products of length-4 subvectors from three groups. The feature maps extracted exhibit a normalized average root mean square error (RMSE) of 0.0243 [Fig. 3(d)]. Additionally, the deviation between the testing and ideal results typically follows a normal distribution. The deviation is primarily attributed to internal coherent interference in the crossbar array, associated with unexpected leakage from MRRs and spectral crosstalk (further details are discussed in Supplementary Note 6).[48]

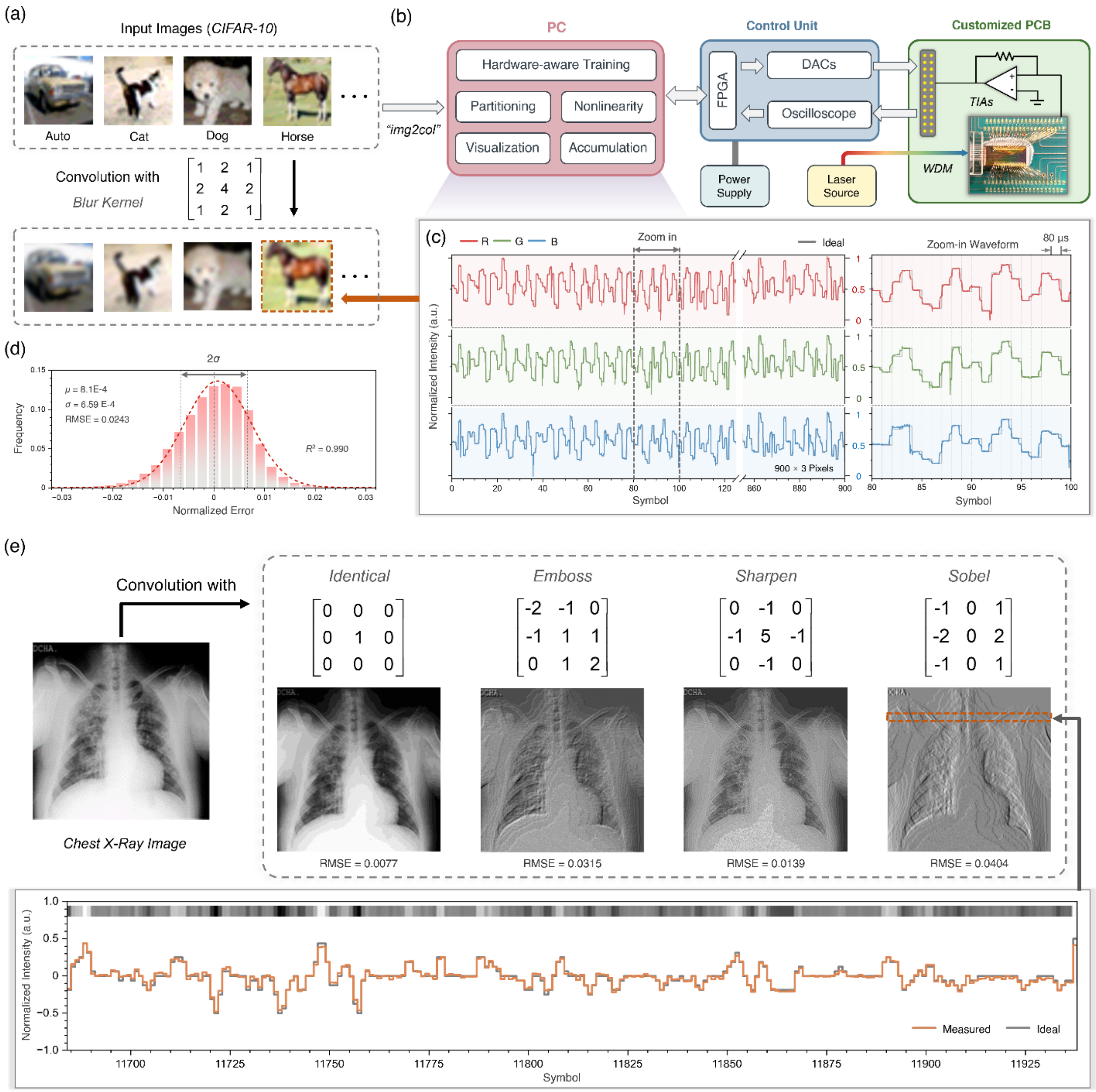

**Fig. 3 Experimental results of the image processing on CirPTC.** (a) Input images from CIFAR-10 dataset processed with a 3×3 blur kernel. (b) Schematic of the experimental setup and test flow. Kernels are first block-circulant extended on PC (for arbitrary kernels) and partitioned into multiple 4×4 blocks. Each block and input vectors are then sent to CirPTC in sequential waveforms, controlled by the FPGA and a multi-channel DAC (c) Ideal and experimental output waveforms for the RGB channels of "horse", and the sampling rate of oscilloscope is 500KHz (40 data points per symbol). (d) Statistical analysis of the error between experimental and ideal feature maps for the CIFAR-10 dataset. (e) X-ray image and the feature maps extracted by four different kernels. Due to the large volume of data, the sampling rate is reduced to 12.5 KHz.

Since the modulators in the CirPTC operate with an amplitude-tuning mechanism, both the weights and inputs are theoretically required to be positive. For ONNs, employing activation functions such as ReLU, sigmoid, or softmax could ensure that the inputs of each layer remain non-negative. The full-range weights can be achieved by two methods. The first method normalizes $\boldsymbol{W}$ to positive values by introducing a reference matrix with all weights set to 1/2 to shift the dynamic range. Specifically, an additional wavelength channel $\lambda^*$ is used to carry input vector elements. After input encoding, this channel is redirected by an added column of optical switches operating at $\lambda^*$. The photocurrent generated from this column serves as a reference and is subtracted from the output signals of the remaining columns, thereby enabling full-range computation with minimal hardware overhead and single-shot input injection. Alternatively, $\boldsymbol{W}$ could be split into two matrices based on the sign of its elements, with each matrix being processed on CirPTC separately. One advantage of post-processing subtraction is that it can automatically eliminate the influence of dark current on the output range. Similarly, both methods can be implemented using either spatial or time-domain multiplexing. The spatial approach requires additional hardware resources, such as balanced photodetectors,[49][50] whereas time-domain multiplexing doubles the processing time and performs subtraction digitally in the post-processing stage.

To further evaluate the capability of CirPTC to process full-range weights, we mapped a chest X-ray (CXR) image (256 × 256 pixels) from the COVID-QU-Ex dataset with a 4-bit resolution,[51] processing with multiple 3×3 kernels. Here, we employ the time-domain multiplexing, wherein the convolution kernels are split into positive and negative parts, each consisting of three 4×4 circulant matrices. Fig. 3(e) shows the extracted features from the CXR image, such as the edges of the human lung highlighted by the vertical Sobel kernel.

## CirPTC-based ONN for classification

In this work, we access the performance of CirPTC on classification tasks over three datasets: a simple CNN is applied to the street view house numbers (SVHN) dataset, while a VGG-style neural network is applied to the CIFAR-10 and COVID-QU-Ex datasets. All convolutional and FC layers are implemented on order-4 CirPTC, while batch normalization (BN), pooling, and nonlinear activation are executed on digital processors. Here, the activation control resolution is set to 4 bits, while the weight precision is configured to 6 bits. The COVID-QU-Ex dataset comprises CXR images from individuals diagnosed with COVID-19, those with non-COVID-19 infections (such as other viral or bacterial pneumonia), and healthy controls. The feature maps shown in Fig. 4(a) are obtained by reshaping the output back into 2D images, which inherently contain inter-kernel dependencies and sparsity arising from structured constrains in the block-circulant weight matrix. For the three-category classification task, the CirPTC-based ONN achieves a classification accuracy of 92.6%, with a sensitivity of 96.3% and a specificity of 98.0% for COVID-19 diagnosis. Furthermore, CirPTC-based ONN experimentally achieved overall accuracies of 80.04% and 88.08% on CIFAR-10 and SVHN tasks, respectively [Fig. 4(b)-(d)].

Additionally, we evaluated the model performance under various configurations, comparing digital and optical approaches, as well as GEMM-based and structure-compressed networks. As shown in Fig. 4(e), the CirPTC-based implementation maintains competitive performance, exhibiting only a 1.41% to 3.65% accuracy drop compared to full-precision GEMM-based digital baselines while achieving up to 74.91% savings in model parameters. Moreover, compared to a digital implementation with circulant compression, the proposed hardware-aware training strategy with DPE ensures that the accuracy drop is kept below 1%. Note that the experiments and simulation indicate that crosstalk and noise from the photonic chip tend to accumulate along the forward path of the network, leading to increased deviation and potentially significant performance degradation. The DPE can mitigate this issue, particularly in ONNs with deep architectures. However, more sophisticated modeling and quantification of on-chip behavior may be necessary for modern large models.

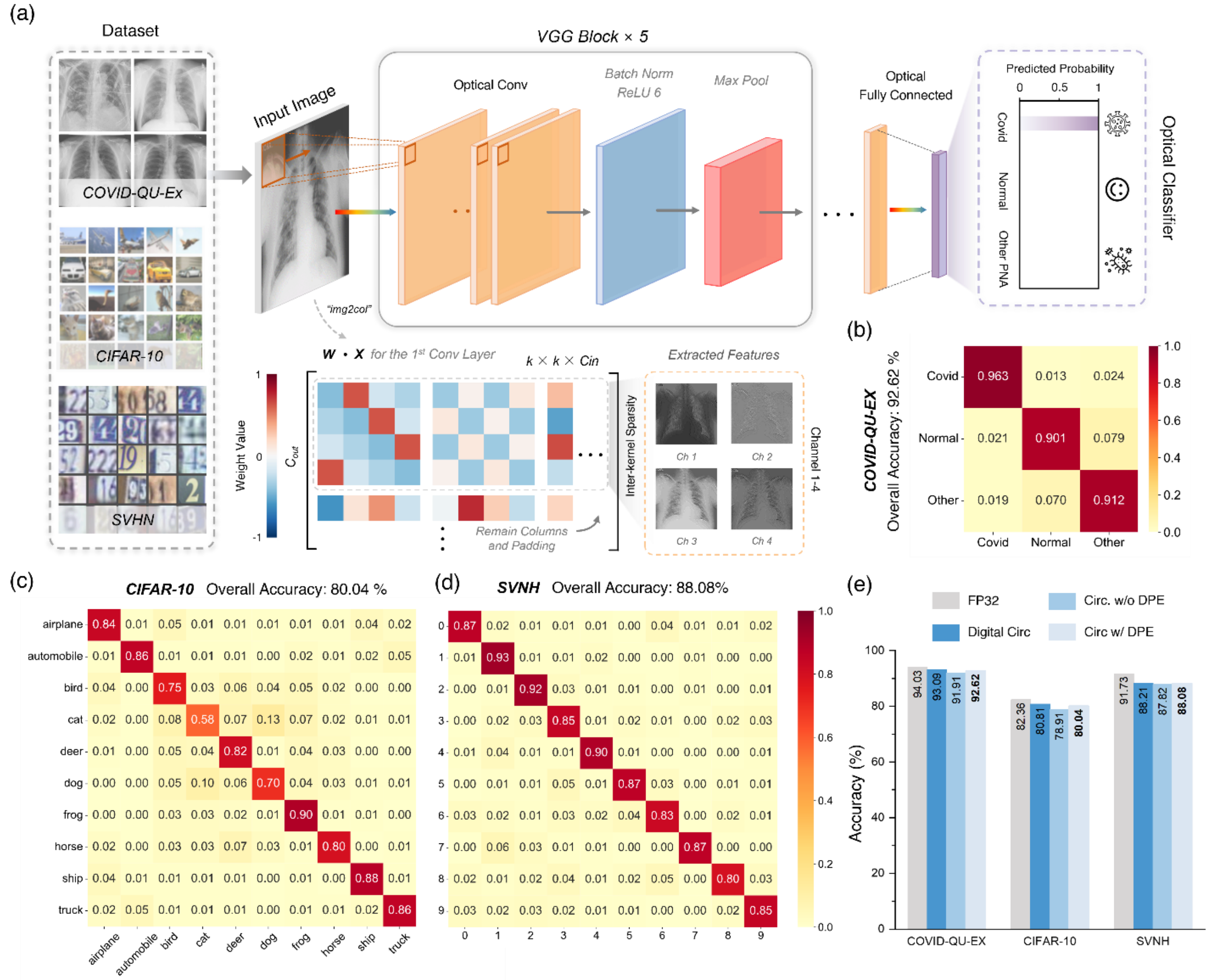

**Fig. 4 Experimental results of the CirPTC-based ONN for image classification.** (a) Structure of StrC-ONNs with input images from multiple datasets, illustrated with a schematic diagram representing a three-category classification task for the COVID-QU-EX dataset. Detailed descriptions of StrC-ONNs implementation are provided in Supplementary Note 7. (b)-(d) Confusion matrices for the three datasets evaluated. (e) Comparison of experimental classification accuracies, with and without DPE-based hardware-aware training, against the simulated accuracies of a digitally structured DNN and a GEMM-based DNN using 32-bit floating-point precision.

## Discussion

### Benchmark analysis

This section evaluates the expected performance of our proposed architecture through numerical analysis, including benchmarks such as chip area, insertion loss, computing speed, power consumption, and scalability. The details of the analysis are discussed in the Supplementary Note 8. Unlike the Von Neumann architecture, PTCs can execute an MVM operation within a single clock cycle. Defining an operation as either the multiplication or addition, the throughput of an $N \times M$ CirPTC can be evaluated by the number of operations per second (OPS), which is quantified as:

$$\mathrm{OPS} = 2MN \cdot f_{op} \tag{2}$$

where $f_{op}$ represents the operational rate. Clearly, enhancing the throughput can be achieved by implementing a larger matrix and increasing $f_{op}$. However, engineering limitations and trade-offs among benchmarks must be considered during the design process. First, $f_{op}$ is dependent on the on-chip devices and the E-O/O-E conversion process. The CirPTC prototype based on thermo-optic devices that we demonstrated experimentally exhibits a tuning speed of tens of KHz.[52] To fully harness the potential of optical computing, operands must be programmed at high speed. To achieve high-speed operation, free-carrier-effect-based MZMs operating in carrier-depletion (reverse-bias *p-n* junction) or carrier-accumulation/metal-oxide-semiconductor capacitor (MOSCAP) modes can replace the thermo-optic MZMs for input encoding in this proof-of-concept demonstration.[25][53][54] Unlike the dynamic input $\boldsymbol{x}$, which updates with each clock cycle, the weights can be shared and remain constant during the inference phase. Therefore, we assume that the modulation speed of thermo-tuned MRRs is sufficient to support the time-domain hardware reuse required for DNNs. Additionally, another factor restricting $f_{op}$ is the delay in PICs. To implement an MVM operation within a single clock cycle, the system clock period $1/f_{op}$ should be no less than the total latency of the CirPTC, which increases linearly with the matrix size.

Alternatively, clock synchronization techniques could enable higher $f_{op}$ while preventing sampling errors.[55]

The computing density, defined by the OPS divided by the chip area, is 4.85 TOPS/mm$^2$ for a 48 × 48 CirPTC operating at 10 GHz. Note that modulators based on the carrier effect typically require larger footprints than thermo-optic devices due to their lower tuning efficiencies and the potential requirement for traveling-wave electrodes to achieve high-speed modulation.

The total power consumption of CirPTC comprises the power to drive lasers, load operands ($\boldsymbol{X}$ and $\boldsymbol{W}$), detect signals, and the static power required to maintain PIC in the operating state. For CirPTC, the static power consumption primarily arises from calibrating MRRs to the desired operating wavelength and maintaining the resonated state, which is negligible when using customized MRRs or post-fabrication nonvolatile phase trimming techniques to correct fabrication variations.[56] Additionally, depletion-mode/MOSCAP or nonvolatile devices can potentially eliminate static power consumption.[24] Based on the references and the experimental results,[52][54] we estimate that each MOSCAP MZM consumes 0.35 pJ per symbol, with each MRR requiring 3 mW to maintain the weight. For output signal detection, the ADC power consumption is 39 mW at 10 GHz and 194 mW at 25 GHz,[57] while the TIA power consumption is 0.65 pJ/bit.[58] Despite the availability of high-speed receivers, the high power consumption of the ADC could be the dominant factor, reducing overall power efficiency [Fig. S17(b)&(f)]. The minimum required laser power must overcome the capacitance and shot noise of the photodetector, as well as to compensate for the insertion loss encountered along the critical path of the PIC.[59] Notably, the insertion loss of CirPTC in the critical path increases linearly with matrix size (Fig. S15), resulting in an exponential increase in laser power. As shown in Fig. S17(e), laser power constitutes 43.14% of the total power when $M = N = 64$, and power efficiency begins to decline. According to our calculations, a 48 × 48 CirPTC configuration achieves the peak power efficiencies of 9.53 TOPS/W. This achieves 1.98× power efficiency higher than those of uncompressed MRR-based crossbar arrays. Beyond the on-chip power consumption, the power required for storing and memory reconfiguration should also be considered. In this work, the memory cost for storing and accessing the weight matrix, as well as the power required to reconfigure active devices, is reduced by a factor of $l$ in order-$l$ CirPTC compared to ONN architectures designed for GEMMs.

In addition to the insertion loss, the scalability of CirPTC is also constrained by the limited density of MRR resonant peaks on the spectrum. Specifically, we encode the weights onto different wavelengths and use WDM techniques to perform $M \times N$ MVM operations across $M$ WDM channels. To avoid errors stemming from spectral crosstalk, the finesse of MRRs needs to accommodate $M$ resonant peaks with permissible overlap, which is evaluated in terms of weight resolution. Based on modeling and numerical analysis, the required $Q$ value for a 6-bit weight resolution is $2.49\times10^5$ when $M = 48$ [Fig. S18]. Although silicon MRRs and microdisk resonators with high $Q$ values above $2\times10^7$ have been widely reported,[60]-[62] fabrication variation and narrow electrical tuning range should be considered. It is also worth noting that, although this work focuses on linear tensor operations, the PIC topology of CirPTC is capable of supporting on-chip nonlinear activation, which are essential for neural network implementation. Specifically, the MZI used for input encoding exhibits a quasi-sigmoidal nonlinear transmission curve. According to our experimental results, the transmission characteristics of the MZI align well with the physical model [Fig. 2(e)]. Therefore, the on-chip nonlinearity can be parameterized and directly integrated into the training framework, eliminating the need for additional electrical-domain post-processing for nonlinear activation.

### Spectral folding

Note that the MRR in the crossbar array functions solely as a wavelength-dependent switch. To further reduce the crossbar array size and enhance CirPTC performance, we propose a spectral folding scaling approach. Specifically, by exploiting its periodicity, a single MRR can redirect signals at different wavelengths across multiple FSRs. This approach enables an $N \times M$ crossbar array to perform the MVM of a BCM with dimensions $M \times (r \cdot N)$ and a length-$r \cdot N$ input vector, where $r$ is the fold number [Fig. S19(a)]. Through spectral folding, the footprint and loss of PIC will be further decreased, thereby improving computing density and power efficiency. Numerical analysis shows that with $r = 4$ and $M = N = 48$, CirPTC achieves a computing density of 5.48 TOPS/mm$^2$ and a power efficiency of 17.13 TOPS/W (3.56× higher than uncompressed MRR-based ONNs), respectively. The significant improvement in power efficiency arises from increased operational throughput without expanding the number of ADCs and TIAs, while the thermal power consumption of the MRRs for weight programming becomes the dominant factor [Fig. S19(b)]. By utilizing depletion-mode/MOSCAP MRRs, this component of power can be potentially eliminated and the power efficiency can be increased to 47.94 TOPS/W. However, this method requires precise spectral alignment of each MRR and more complex control schemes, thereby necessitating advanced fabrication techniques capable of achieving the tolerances required to counter device-to-device variation. The residual misalignments can be further compensated by the available programming range of the weight banks or by adjustments to laser power. Details of the spectral-folding technique are provided in Supplementary Note 8, and Table S6 compares our approach with state-of-the-art optical and electrical computing architectures.

## Conclusion

In this work, we propose CirPTC, a scalable photonic-electric hybrid AI accelerator with a hardware-efficient ONN architecture using a structured compression technique. We experimentally demonstrate on-chip convolution processing of large-scale images. Then, the StrC-ONN architecture is implemented on the order-4 CirPTC for image classification tasks. By reasonable scaling and spectral folding approach, the proposed design achieves 3.56× higher power efficiency compared to uncompressed MRR-based ONNs, while using ~25%-35% of model parameters, active optical components, and memory usage. Additionally, we employ a hardware-aware training framework incorporating the DPE, which efficiently models the on-chip behavior of CirPTC, accounting for nonidealities such as inherent crosstalk and noise, thereby boosting the model robustness. Notably, CirPTC-based ONNs with circulant structured compression achieve comparable performance across multiple datasets to full-precision digital GEMM-based DNNs, demonstrating negligible loss in accuracy. Furthermore, the compression strategy and the DPE-based training framework can be extended to existing PICs, enhancing their hardware and power efficiency. These findings offer a novel route to overcoming the bottlenecks of optical computing, thus paving the way for next-generation high-performance AI accelerators in the post-Moore era.

## Funding

Multidisciplinary University Research Initiative (MURI) Program (FA9550-17-1-0071); Air Force Office of Scientific Research (AFOSR) (FA9550-23-1-0452).

## Disclosures

The authors declare no conflicts of interest.

## Data availability

Data underlying the results presented in this paper are available from the authors upon reasonable request.

## Supplemental document.

See Supplementary for supporting content.